\documentclass[a4paper]{article}

\usepackage{INTERSPEECH2022}
\usepackage{multirow,amsmath,bm,amssymb,cite}

\title{Speaker Embedding-aware Neural Diarization: an Efficient Framework for Overlapping Speech Diarization in Meeting Scenarios}
\name{Zhihao Du, Shiliang Zhang, Siqi Zheng, Zhijie Yan}
\address{Speech Lab, Alibaba Group, China}
\email{\{neo.dzh,sly.zsl\}@alibaba-inc.com}
\begin{document}

\maketitle
\begin{abstract}
  Overlapping speech diarization has been traditionally treated as a multi-label classification problem.
  In this paper, we reformulate this task as a single-label prediction problem by encoding multiple binary labels into a single label with the power set, which represents the possible combinations of target speakers.
  This formulation has two benefits. First, the overlaps of target speakers are explicitly modeled. Second, threshold selection is no longer needed.
  Through this formulation, we propose the speaker embedding-aware neural diarization (SEND) framework, where a speech encoder, a speaker encoder, two similarity scorers, and a post-processing network are jointly optimized to predict the encoded labels according to the similarities between speech features and speaker embeddings.
  Experimental results show that SEND has a stable learning process and can be trained on highly overlapped data without extra initialization.
  More importantly, our method achieves the state-of-the-art performance in real meeting scenarios with fewer model parameters and lower computational complexity.
\end{abstract}
\noindent\textbf{Index Terms}: power-set encoding, overlapping speech, speaker diarization

\section{Introduction}
\label{sec:intro}
Speaker diarization aims at answering the question ``who spoke when.''
It is an important technique for speech applications in the real world, such as speaker-attributed automatic speech recognition (SA-ASR) \cite{CarlettaABFGHKKKKLLLMPRW05,BarkerWVT18}.

Classical clustering-based diarization systems comprise three individual parts, including speech segmentation, embedding extraction, and unsupervised clustering algorithm.
In this approach, an audio recording is first split into several segments by voice activity detection (VAD). 
Then, speaker embeddings, such as i-vector \cite{DehakKDDO11}, d-vector \cite{WanWPL18}, and x-vector \cite{SnyderGSPK18}, are extracted from each segment.
Finally, speaker embeddings are partitioned into clusters by unsupervised clustering algorithms, such as k-means \cite{DimitriadisF17}, spectral clustering \cite{NingLTH06}, agglomerative hierarchical clustering (AHC) \cite{Garcia-RomeroSS17}, and Leiden community detection \cite{ZhengSQ2022}.
Recently, the variational Bayesian hidden Markov model has been introduced to improve the clustering results of x-vector sequences (VBx) \cite{DiezBWRC19}. Although VBx achieves impressive performance on several datasets \cite{CastaldoCDLV08, RyantCCCDGL19}, it has limited ability to handle overlapping speech due to the speaker-homogeneous assumption in each segment.
Besides, clustering-based algorithms do not aim at minimize the diarization errors directly, which are performed in an unsupervised manner.

End-to-end neural diarization (EEND) \cite{FujitaKHNW19} is introduced to deal with overlapping speech by minimizing the utterance-level permutation-invariant training (uPIT) loss \cite{KolbaekYTJ17}.
Subsequently, the self-attention mechanism \cite{FujitaKHXNW19} and conformer architecture \cite{Liu2021Conformer} are employed to further improve the performance of EEND.
To deal with an unknown number of speakers, the encoder-decoder based attractor is involved into EEND \cite{HoriguchiF0XN20}.
However, EEND has trouble handling a large number of speakers at training stage due to the label permutation problem.

Target-speaker voice activity detection (TSVAD) is another approach to handle overlapping speech \cite{MedennikovKPKKS20}, which can avoid the label permutation problem by giving the embeddings of target speakers. In \cite{Weiqing2022}, TSVAD suffers from unstable training, which can not converge on a highly overlapped dataset without extra initialization.
In both TSVAD and EEND, overlapping speech diarization is formulated as a \emph{multi-label} classification problem where target speakers are detected independently.
This formulation ignores the correlation between different speakers, which is not desired in real meeting scenarios.
Besides, the final diarization results of TSVAD and EEND heavily depend on the threshold selection.

In this paper, we attempt to solve the problems by reformulating overlapping speech diarization from a \emph{multi-label} classification problem to a \emph{single-label} prediction problem.
Through this formulation, we propose a novel framework named speaker embedding-aware neural diarization (SEND).
In SEND, multiple binary labels of each frame are encoded into a single label using the power set of target speakers. Each encoded label represents a possible combination of different speakers. 
In this way, SEND can model the overlaps of different speakers explicitly, and threshold selection is no longer needed.
As a result, SEND achieves a better diarization performance than the state-of-the-art methods in real meeting scenarios with fewer parameters and less computational complexity.
Another benefit of SEND is the improved stability of training process.
Unlike TSVAD, our model can be trained on a complicated dataset from scratch without extra initialization.

\section{Proposed framework}
\subsection{Speaker embedding-aware neural diarization}
\begin{figure}[t!]
	\centering
	\includegraphics[width=0.80\linewidth]{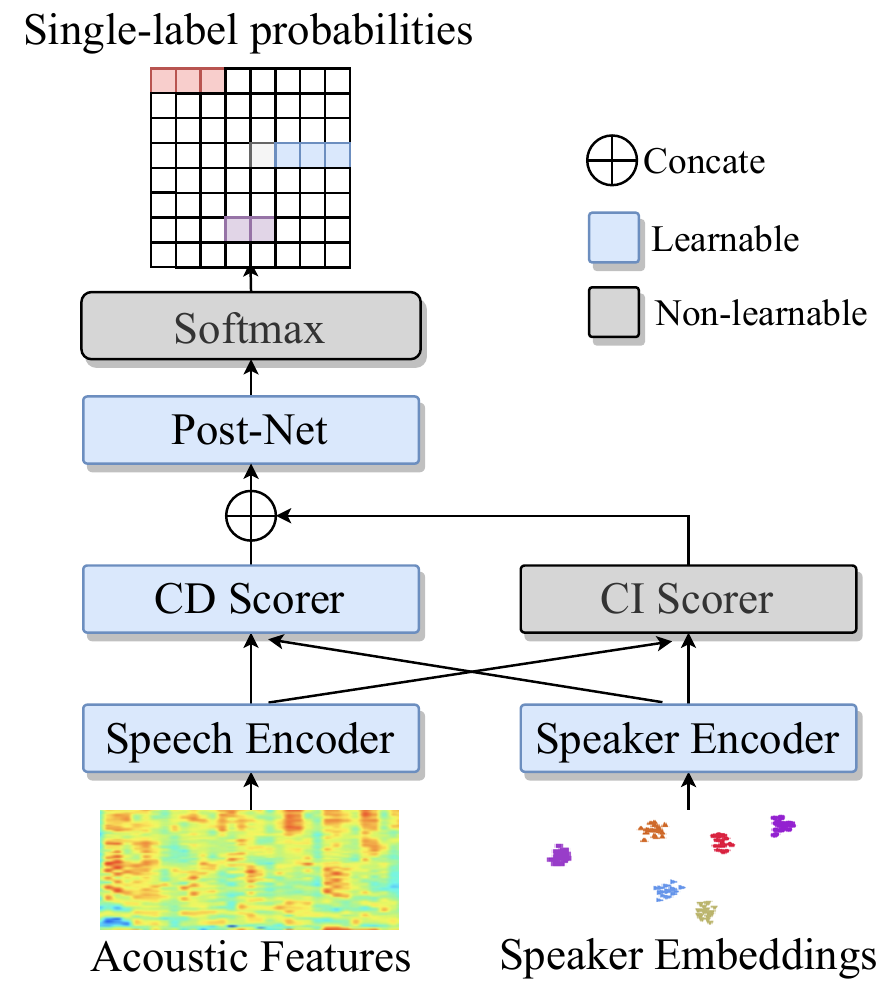}
	\caption{Schematic diagram of the proposed SEND framework.}
	\label{fig:send}
\end{figure}

The schematic diagram of SEND is given in Figure \ref{fig:send}.
At the beginning of our system, acoustic features and speaker embeddings are encoded by the speech and speaker encoders.
Then, the outputs of encoders are fed to the context-independent (CI) and context-dependent (CD) scorers simultaneously.
Subsequently, CI and CD scores are concatenated and provided to the post-processing network (post-net).
After a softmax activation function, the probabilities of power-set encoded (PSE) labels are predicted.
\subsubsection{Deep feed-forward sequential memory network}
We employ the deep feed-forward sequential memory networks (FSMNs) \cite{zhang2015fsmn} as the basic architecture of the speech encoder and post-net.
Deep FSMNs are widely used in sequence modeling tasks, especially in the speech processing community \cite{BiLZLY18, ZhangLYD18,ZhangLLL19a}.
In each FSMN layer, there is a memory block defined as follows:
\begin{equation}
\begin{split}
	&\tilde{\mathbf{h}}^\ell_t =\sum_{i=0}^{L_l}\mathbf{a}_i^\ell\odot \mathbf{h}^\ell_{t-i\cdot d^\ell} + \sum_{j=0}^{L_r}\mathbf{c}_j^\ell\odot \mathbf{h}^\ell_{t+j\cdot d^\ell} \\
	&\mathbf{h}^{\ell+1}_t =\sigma(\mathbf{W}^\ell\mathbf{h}_t^\ell + \tilde{\mathbf{W}}^\ell\tilde{\mathbf{h}}^\ell_t + \mathbf{b}^\ell)
\end{split}
\end{equation}
where $\odot$ denotes element-wise multiplication, and $\sigma$ represents an activation function.
$\mathbf{h}^\ell_t$ denotes the hidden states at time step $t$ in layer $\ell$. 
$L_l$ and $L_r$ represent the window sizes of history and future frames, respectively.
$\mathbf{a}_i^\ell$ and $\mathbf{c}_j^\ell$ are learnable vectors to aggregate the information at time steps $t-i$ and $t+j$. $d^\ell$ denotes the stride factor for layer $\ell$.
By setting stride factors properly, the receptive field of FSMN can increase exponentially, and the redundancy of adjacent frames is removed, resulting in the modeling ability of long-term dependency.


\subsubsection{Context-independent and context-dependent scorers}
In the proposed model, there are two scorers to compute the similarities between speech features and speaker embeddings according to the context-independent and context-dependent information.
Given encoded speech features $\mathbf{H}=\{\mathbf{h}_t|t=1,\dots,T\}$ and speaker embeddings $\mathbf{E}=\{\mathbf{e}_n|n=1,\dots,N\}$, the context-independent (CI) score $S^{CI}_{t,n}$ is derived from the dot product of $\mathbf{h}_t$ and $\mathbf{e}_n$:
\begin{equation}
	S^{CI}_{t,n}=<\mathbf{h}_t, \mathbf{e}_n>
\end{equation}
We also try to utilize the cosine similarity as the context-independent score, but the model can not converge for a long time.
While context-independent scores only consider current speech encodings, the contextual information of different speakers is also crucial for identifying the activated speaker from others.
Therefore, we further employ a context-dependent (CD) scorer in SEND. The CD score $S^{CD}_{t,n}$ is defined as follows:
\begin{equation}
	S^{CD}_{t,n} = f(\mathbf{h}_t, \mathbf{e}_n;\mathbf{H},\Theta)
\end{equation}
where $f$ is a context-aware function, such as the bidirectional long short-term memory (BiLSTM) \cite{hochreiter1997long} or self-attention based networks (SAN) \cite{VaswaniSPUJGKP17}.
$\Theta$ represents the learnable parameters of function $f$.
\begin{figure}[t!]
	\centering
	\includegraphics[width=0.9\linewidth]{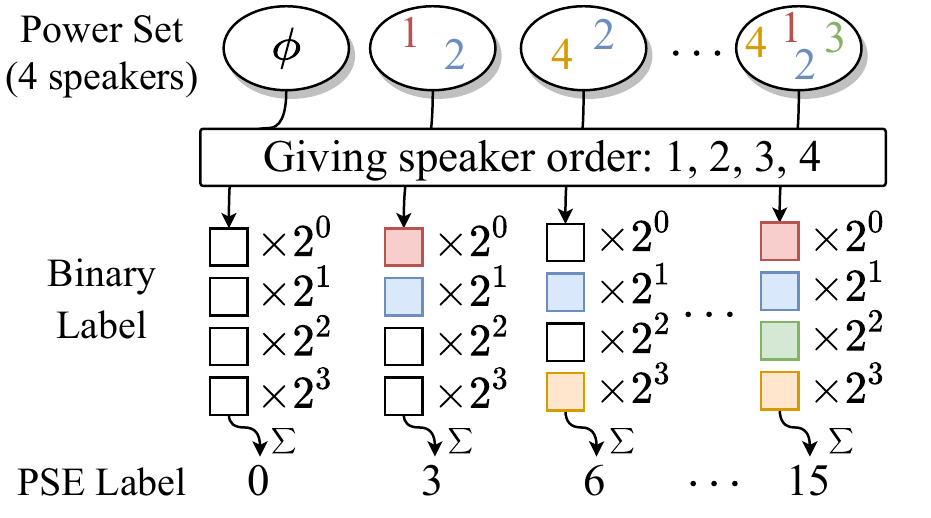}
	\caption{Illustration of power set encoding.}
	\label{fig:pse}
\end{figure}
\subsubsection{Power-set encoding for overlapping speech diarization}
In previous studies, overlapping speech diarization is always formulated as a multi-label classification problem by treating each speaker as a category, where the binary label $y_{t,n}$ indicates whether speaker $n$ talks at frame $t$.
In this paper, we reformulate overlapping speech diarization as a single-label prediction problem through a power set.
Given $N$ speakers $\{1,2,\dots,N\}$, their power set (PS) is defined as follows:
\begin{equation}
	\begin{split}
		\mathrm{PS}(N) &=\{A|A\subseteq \{1,2,\dots,N\}\} \\
		&=\{\phi,\{1\},\{2\},\dots,\{1,2,n,\dots\}, \dots \}         
	\end{split}
\end{equation}
where $\phi$ means the empty set. 
From the definition, we can see that each element of PS represents a combination of speakers, and the power set contains all possible combinations. 
Therefore, if we treat the elements of PS as classification categories, an overlapping speech frame can be uniquely assigned with a single label.
In this paper, we employ a simple approach to encode the PS elements, where a power-set encoded (PSE) label $\tilde{y}_t$ is obtained by treating the binary label $y_{t,n}$ as an indicator variable:
\begin{equation}
	\tilde{y}_t= \sum_{n=1}^{N}{y_{t,n}\cdot 2^{n-1}}
\end{equation}
The procedure of PSE is illustrated in Figure \ref{fig:pse}, where $\sum$ means the summation operator.
By applying power-set encoding on $N$ speakers, we can get $2^N$ categories, which may be impractical for a large number of speakers. 
Fortunately, the maximum number of overlapping speakers $K$ is always small (e.g., two, three or four at most) in real-world applications. Therefore, the number of reasonable categories can be reduced to:
\begin{equation}
	\mathcal{C}(K,N) = \sum_{k=0}^{K}{N \choose k} = \sum_{k=0}^{K}\frac{N!}{k!(N-k)!}
\end{equation}
Finally, overlapping speech diarization is reformulated as a single-label prediction problem with $\mathcal{C}(K,N)$ categories.

\subsection{Deep speaker embedding extraction}
\subsubsection{Speaker embedding model}
We employ the time-delay neural network (TDNN) as our speaker embedding model with the same architecture described in \cite{SnyderGPK17}.
The frame aggregation layer is based on statistic pooling, and the dimension of speaker embeddings is 512.
The additive angular softmax \cite{DengGXZ19} with a margin of 0.25 is adopted to train the speaker embedding model.
The input feature is 80-dim log-compressed Mel-filterbank energies (log-fbank) with a frame length of 25ms and a frame shift of 10ms.

\subsubsection{Data augmentation}
To enrich training samples, we perform data augmentation with the MUSAN noise dataset \cite{musan2015} and simulated room impulse responses (RIRs) \cite{KoPPSK17}.
We first perform the amplification and tempo perturbation (change audio playback speed but do not change its pitch) on speech.
Then, 40,000 simulated RIRs from small and medium rooms are used for reverberation.
Finally, the reverberated signals are mixed with background noises at the speech-to-noise rates (SNRs) of 0, 5, 10, and 15 dB.

\section{Experimental settings and results}
\subsection{Experimental settings}
\subsubsection{Experimental settings of speaker embedding model}
We conduct experiments on the CN-Celeb \cite{FanKLLCCZZCW20} corpus and the first channel data from the AliMeeting dataset \cite{FanYu2022}\footnote{Available at http://openslr.org/119/}, which contains 240 meetings (about 120 hours) recorded in real meeting scenarios. 
The speech overlap ratio of AliMeeting is about 34\%$\sim$42\%. In each meeting, the number of speakers ranges from 2 to 4.
Since the AliMeeting dataset does not provide single-speaker utterances, we select non-overlapping speech from entire meetings according to ground-truth transcriptions, where segments shorter than two seconds are dropped.
We build a trial set from the AliMeeting evaluation set to determine model parameters and the number of training steps. 
Another trial set is built from the AliMeeting test set to evaluate the performance of the speaker embedding model. Table \ref{tab:eer} reports the equal error-rate (EER) results and the minimum of the normalized detection cost function (mDCF) at $P_{\text{Target}}=0.01$.

\linespread{1.1}
\begin{table}[t!]
	\caption{The performance of speaker embedding model.}
	\label{tab:eer}
	\centering
	\setlength{\tabcolsep}{1.4mm}{
	\begin{tabular}{lccccc}
		\toprule
		\multirow{2}{*}{\textbf{Enroll}} & \multicolumn{2}{c}{\textbf{Evaluation set}} & & \multicolumn{2}{c}{\textbf{Test set}} \\
		\cline{2-3}\cline{5-6}
		& \textbf{EER(\%)} & \textbf{mDCF}$_{0.01}$ && \textbf{EER}(\%) & \textbf{mDCF}$_{0.01}$ \\
		\midrule
		\linespread{1.0}
		50s & 1.634 & 0.0712 && 2.565 & 0.1307 \\
		250s & 1.299 & 0.0522 && 1.842 & 0.0994 \\
		\bottomrule
	\end{tabular}}
	\vspace{-0.4cm}
\end{table}
\subsubsection{Dataset for diarization model}
Diarization models are first trained on a simulated dataset and then fine-tuned with the AliMeeting corpus \cite{FanYu2022}.
We adopt a similar simulation process as described in \cite{Weiqing2022}. First, all non-overlapping speech segments are extracted from the AliMeeting training set for each speaker.
Then, transcript files are converted to frame-level labels, and silence regions are removed.
To simulate a training sample, we choose a label segment and fill the activated region with non-overlapping speech segments. Each simulated sample has a fixed length of 16s.

For the AliMeeting training set, we first remove silence regions according to the transcript files. Then, the left segments are concatenated and split into chunks with a fixed length of 16s and a shift of 4s. The evaluation and test sets of the AliMeeting corpus are processed in the same manner as the training set.
\subsubsection{Experimental settings of SEND}
A multi-layer perceptron (MLP) is adopted as the speaker encoder, which consists of three layers and 512 hidden units in each layer.
Deep FSMNs are employed as the speech encoder and post-net.
While there are eight layers in the speech encoder, the post-net comprises six FSMN layers.
There are 512 memory units in each FSMN layer, and the stride factors of the first four layers are 1, 2, 4, 8.
As for the context-dependent scorer, we employ a four-layer transformer encoder with four attention heads and 512 hidden units in each layer.

We first train the models on simulated data for 200,000 steps with a learning rate of 1.0 and warm-up steps of 10,000.
Next, we fine-tune models on the AliMeeting training set for 50,000 steps with a constant learning rate of $10^{-5}$. Three best models with the least evaluation errors are averaged as the final model for evaluation and test.
The softmax cross-entropy (CE) loss and Adam optimizer \cite{KingmaB14} are used to train the models.
Diarization results are smoothed by a median filter with a window size of 83 (about 0.83s).

\subsubsection{Baselines}
The VBx algorithm \cite{landini2022bayesian} and TSVAD model \cite{MedennikovKPKKS20} are employed as baselines.
We reuse the official code released by the M2MeT challenge \cite{FanYu2022} to implement the VBx algorithm.
As for the TSVAD model, we adopt a similar architecture and training process as described in \cite{Weiqing2022}. 
The TSVAD model is trained to minimize the binary cross-entropy (BCE) loss.
In both baselines and SEND, the maximum speaker number $N$ is set to four as well as the overlapped speaker number $K$.
Speaker embeddings are extracted from non-overlapping speech segments according to transcript files.
We employ the commonly-used diarization error rate (DER) as the evaluation metric\footnote{Codes are available at https://github.com/nryant/dscore}.
Since the activation threshold severely affects the final diarization performance of TSVAD, we report the averaged DER among three best results for both evaluation and test sets with the search grid of 0.1. Note that such TSVAD model is a strong baseline, which achieves the best performance in the M2MeT challenge \cite{Yu2022Summary}.

\subsection{Results}
\begin{table}[t!]
	\caption{The diarization error rates (\%) of different methods. Model parameters and floating-point operations (FLOPs) are counted in millions (M).}
	\label{tab:der}
	\centering
	\setlength{\tabcolsep}{1.0mm}{
		\begin{tabular}{lcccccc}
			\toprule
			\textbf{Model} & \textbf{Training set} & \textbf{Eval.} & \textbf{Test} & \textbf{Avg.} & \textbf{Size} & \textbf{FLOPs} \\
			\midrule
			VBx & sim.+real & 15.24 & 15.60 & 15.42 & - & -\\
			\hline
			\multirow{2}{*}{TSVAD} & sim. & 5.91 & 8.70 & 7.31 & \multirow{2}{*}{21.16} & \multirow{2}{*}{60.80} \\
			& sim.+real & 4.36 & \textbf{5.59} & 4.98 & & \\
			\hline
			\multirow{2}{*}{SEND} & sim. & 5.22 & 8.11 & 6.67 & \multirow{2}{*}{18.42} & \multirow{2}{*}{36.73} \\ 
			& sim.+real & \textbf{3.96} & 5.80 & \textbf{4.88} & & \\
			\bottomrule
	\end{tabular}}
	\vspace{-0.3cm}
\end{table}
\subsubsection{Comparison with baselines}
As shown in Table \ref{tab:der}, supervised methods (TSVAD and SEND) significantly outperform the clustering-based algorithm, VBx, in terms of DER.
That is because the averaged speech overlap ratio of AliMeeting is very high, and the VBx algorithm can not deal with overlapping speech segments.
We find that supervised methods have a performance gap among the evaluation and test sets.
This is mainly because we use the evaluation set to determinate the model parameters, and there is a mismatch between the overlap ratios of evaluation and test sets (34.20\% vs. 42.80\%).
Compared with TSVAD, our SEND achieves better performance on the evaluation set, no matter the model is fine-tuned with real meeting data or not.
On the test set, our method outperforms TSVAD when models are only trained with the simulated data.
This indicates that our method has better generalization ability on unseen data.
When the real data is adopted, our method achieves a lower averaged DER among the evaluation and test sets than baselines.

In addition, SEND has fewer trainable parameters than TSVAD, as shown in the column ``Size'' of Table \ref{tab:der}. 
This is mainly due to the use of shared weights in the memory blocks of FSMN.
With a higher parameter efficiency, SEND is easier to train than TSVAD. 
Moreover, SEND needs much fewer floating-point operations (FLOPs) than TSVAD, which can achieve much higher inference efficiency and lower latency.
\subsubsection{Impact of power-set encoding}
We evaluate the impact of power-set encoding on both SEND and TSVAD. The results are given in Table \ref{tab:pse}.
From the table, we can see that, without PSE, SEND achieves a similar performance as TSVAD in terms of averaged DER.
When the PSE is adopted, the diarization performance of SEND and TSVAD is improved consistently.
In addition, we evaluate the effect of maximum overlapping speaker number $K$ on SEND.
Experimental results show that a larger $K$ can provide a better diarization performance.
This is because a larger $K$ means more encoded speaker combinations, and the labels of overlapping speech segments can be more accurate.

\begin{table}[t!]
	\caption{The impact of power-set encoding on different models in terms of DER(\%).}
	\label{tab:pse}
	\centering
	\setlength{\tabcolsep}{3.5mm}{
		\begin{tabular}{lccccc}
			\toprule
			\textbf{Model} & \textbf{PSE} & \textbf{K} & \textbf{Eval.} & \textbf{Test} & \textbf{Avg.} \\
			\midrule
			TSVAD & $\times$ & - & 4.36 & 5.59 & 4.98 \\
			TSVAD & $\surd$ & 4 & 4.22 & \textbf{5.55} & \textbf{4.89} \\
			SEND & $\times$ & - & 4.13 & 5.85 & 4.99 \\
			SEND & $\surd$  & 2 & 5.17 & 7.63 & 6.40  \\
			SEND & $\surd$  & 3 & 4.22 & 5.79 & 5.01 \\
			SEND & $\surd$ & 4 & \textbf{3.96} & 5.80 & \textbf{4.88} \\
			\bottomrule
	\end{tabular}}
\end{table}

\subsubsection{Ablation study}
\begin{table}[t!]
	\caption{The ablation study on SEND in terms of DER(\%).}
	\label{tab:abla}
	\centering
	\setlength{\tabcolsep}{1.2mm}{
		\begin{tabular}{lccccc}
			\toprule
			\textbf{Model} & \textbf{Exp. Stride} & \textbf{CD} & \textbf{Eval.} & \textbf{Test} & \textbf{Avg.} \\
			\midrule
			SEND & $\surd$ & $\surd$ & 3.96 & 5.80 & 4.88 \\
			~~w/o CD Scorer & $\surd$ & $\times$ & 4.37 & 6.69 & 5.53 \\
			~~~~w/o Exp. Stride & $\times$ & $\times$ & 5.97 & 8.68 & 7.33 \\
			\bottomrule
	\end{tabular}}
	\vspace{-0.3cm}
\end{table}
We conduct an ablation study to evaluate the impact of components in SEND, and the results are shown in Table \ref{tab:abla}.
From the table, we can see that, removing CD scorer from SEND leads to a 13.32\% relative performance degradation in terms of the averaged DER among the evaluation and test sets.
This is mainly because the CD scorer can provide more discrimination between speakers by utilizing the contextual information of entire utterance.
Further replacing the exponential (Exp.) stride factors with linear ones leads another 32.55\% relative performance degradation, which indicates that the modeling ability of long-term dependency is crucial for speaker diarization tasks.

\begin{figure}[t!]
	\begin{minipage}[b]{1.0\linewidth}
		\centering
		\centerline{\includegraphics[width=8cm]{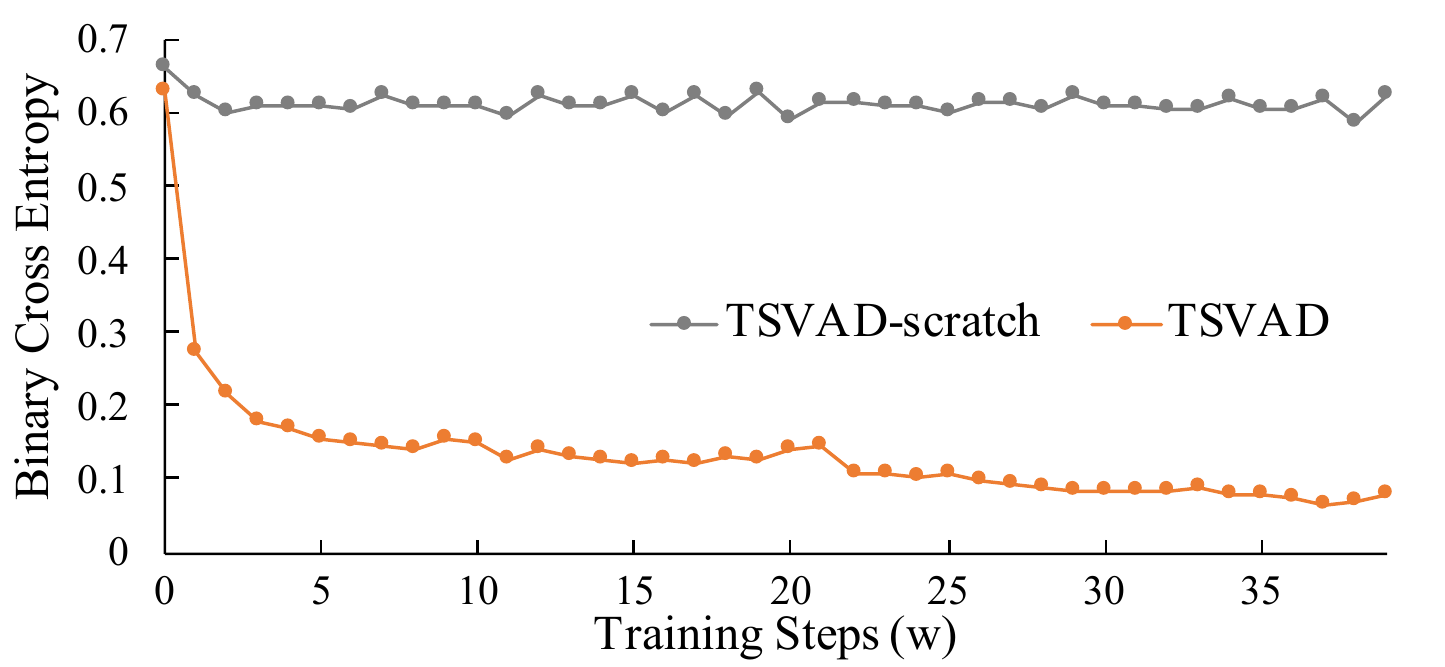}}
		\centerline{(a) TSVAD}\medskip
	\end{minipage}
	\hfill
	\begin{minipage}[b]{1.0\linewidth}
		\centering
		\centerline{\includegraphics[width=8cm]{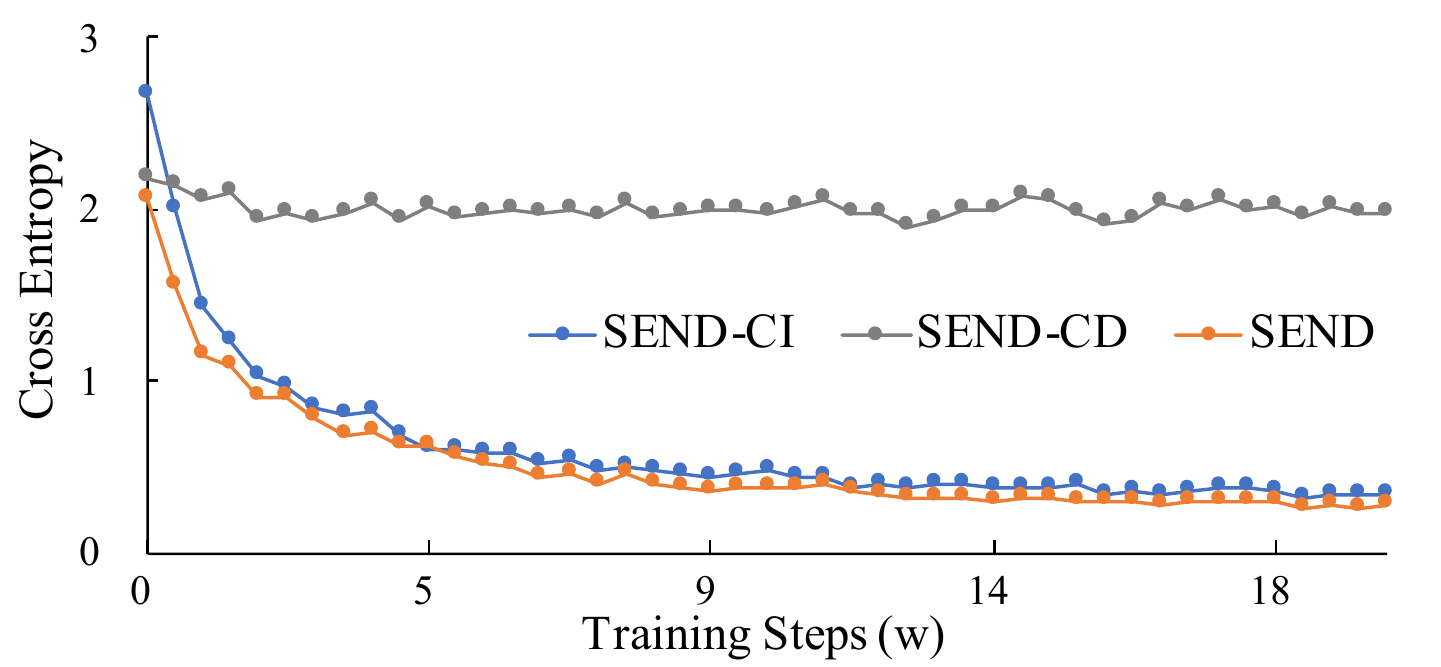}}
		\centerline{(b) SEND}\medskip
	\end{minipage}
	\caption{Cross entropy losses over training steps for (a) TSVAD and (b) SEND on the training set.}
	\label{fig:stab}
	\vspace{-0.3cm}
\end{figure}
\subsubsection{Training stability of SEND and TSVAD}
\label{sec:train}
In this section, we compare the training stability of TSVAD and SEND. As shown in Figure \ref{fig:stab}(a), the TSVAD model can not converge when it is trained from scratch. To make it converge, we have to initialize the front-end model of TSVAD with the pre-trained speaker embedding model and only train the back-end model. After it converges, we jointly train the whole model with simulated data and then fine-tune it on the real dataset. 

Compared with TSVAD, the training process of SEND is more straightforward.
As shown in Figure \ref{fig:stab}(b), SEND can converge when trained from scratch.
To figure out the reason of stable training process, we also demonstrate models with only CI scorer (SEND-CI) and CD scorer (SEND-CD) in Figure \ref{fig:stab}(b).
From the figure, we find that the CI scorer is crucial to the training stability.
When the CI scorer is removed from SEND, the model can not converge for a long time.
On the contrary, if we remove the CD scorer from SEND, the averaged diarization error rate on the evaluation and test sets would increase from 4.88\% to 5.53\%, resulting in about 13.32\% relative degradation (seen in Table \ref{tab:abla}).
 
\section{Conclusions}
In this paper, we reformulate overlapping speech diarization from a \emph{multi-label} classification problem into a \emph{single-label} prediction problem by using the PSE, which can avoid the threshold selection and explicitly model all possible overlap situations of target speakers. Through this formation, a novel framework, SEND, is proposed to predict the power-set encoded labels according to the CI and CD scores between speech features and speaker embeddings.
Benefiting from the CI scorer, SEND has a more stable learning process.
Combining with CD scorer and PSE, SEND achieves the state-of-the-art performance on a real meeting dataset with fewer model parameters and lower computational complexity.

\bibliographystyle{IEEEtran}

\bibliography{mybib}

\end{document}